\documentclass[8.5pt,twoside,twocolumn]{article}
\usepackage[super,sort&compress,comma]{natbib}
\usepackage{mhchem}
\usepackage{times,mathptmx}
\usepackage{sectsty}
\usepackage{balance} 

\usepackage{graphicx} 
\usepackage{lastpage}
\usepackage[format=plain,justification=raggedright,singlelinecheck=false,font=small,labelfont=bf,labelsep=space]{caption} 
\usepackage{fancyhdr}

\begin{document}

\thispagestyle{plain}
\fancypagestyle{plain}{
\renewcommand{\headrulewidth}{1pt}}
\renewcommand{\thefootnote}{\fnsymbol{footnote}}
\renewcommand\footnoterule{\vspace*{1pt}%
\hrule width 3.4in height 0.4pt \vspace*{5pt}} 
\setcounter{secnumdepth}{5}

\makeatletter 
\def\subsubsection{\@startsection{subsubsection}{3}{10pt}{-1.25ex plus -1ex minus -.1ex}{0ex plus 0ex}{\normalsize\bf}} 
\def\paragraph{\@startsection{paragraph}{4}{10pt}{-1.25ex plus -1ex minus -.1ex}{0ex plus 0ex}{\normalsize\textit}} 
\renewcommand\@biblabel[1]{#1}            
\renewcommand\@makefntext[1]%
{\noindent\makebox[0pt][r]{\@thefnmark\,}#1}
\makeatother 
\renewcommand{\figurename}{\small{Fig.}~}
\sectionfont{\large}
\subsectionfont{\normalsize} 

\renewcommand{\headrulewidth}{1pt} 
\renewcommand{\footrulewidth}{1pt}
\setlength{\arrayrulewidth}{1pt}
\setlength{\columnsep}{6.5mm}
\setlength\bibsep{1pt}

\twocolumn[
  \begin{@twocolumnfalse}
\noindent\LARGE{\textbf{Simulating the pitch sensitivity of twisted nematics of patchy rods}} 
\vspace{0.6cm}

\noindent\large{\textbf{\v{S}t\v{e}p\'{a}n R\r{u}\v{z}i\v{c}ka and  Henricus H.\ Wensink}}\vspace{0.5cm}



\noindent\textit{\small{\textbf{Received Xth XXXXXXXXXX 20XX, Accepted Xth XXXXXXXXX 20XX\newline
First published on the web Xth XXXXXXXXXX 200X}}}

\noindent \textbf{\small{DOI: 10.1039/b000000x}}
\vspace{0.6cm}

\noindent \normalsize{
Stiff, elongated biomolecules such as filamentous viruses, DNA  or cellulose nanocrystals are known to
form liquid crystals often exhibiting a helical supramolecular organization.
Little is known about the microscopic origin, size and handedness of the helical pitch in these, so-called cholesteric phases.  Experimental observations in chiral lyotropics suggest that long-ranged chiral forces of electrostatic origin acting between the mesogens are responsible for such organization. 
Using large-scale computer simulation we study the sensitivity of the pitch imparted by soft microscopic helices and confirm that the helical sense is sensitive to a change of packing fraction, magnitude of the molecular pitch and amplitude of the chiral interactions. In particular, we find evidence that  the cholesteric helix sense may change spontaneously upon variation of particle density, at {\em fixed} molecular chirality. These pitch inversions have been reported in recent theoretical studies but simulation evidence remains elusive. We rationalize these sudden changes in the supramolecular helical symmetry on the basis of detailed measurements of the mean-torque
 generated by the twisting of the helices. 
The simulation methodology employed does not require confining the twisted nematic in a slab geometry and allows for a simultaneous measurement of the pitch and the twist elastic constant.  We find that the twist elastic constant increases almost linearly with density suggesting that twisted nematic shows no signs of anomalous stiffening due to pre-smectic fluctuations at higher packing fraction.
}

\vspace{0.5cm}
 \end{@twocolumnfalse}
  ]

\section{Introduction}

Chirality is generally the absence of symmetry and plays a key role in the physics of liquid crystals \cite{harris1999}.
A molecule is termed chiral when, if reflected,
any combination of rotational and translational operations
fails to bring the molecule back to its original shape.
A rigid linear achiral molecule is often represented by a uniaxial rod such as a hard spherocylinder.
A system of many rods is a basic model for lyotropic liquid crystalline mesophases.
The first such phase is the \textit{nematic} phase,
where rods align along a similar direction given by the \textit{director} vector $\vec{n}$.
Upon increasing the density,
the nematic phase transforms into a lamellar structure  consisting of layers of aligned rods with liquid-like intralayer structure called the \textit{smectic} phase.
The reflectional and uniaxial symmetry of the rods can be broken 
by the presence of interaction sites,
representing specific chemical groups or charges residing on the particle surface,  that render the intermolecular potential chiral.
The chirality of the `patchy' rod is referred to as \textit{microscopic}
and is generally weak compared to the achiral steric interactions between the rods
which in turn stabilize nematic order.
The microscopic chirality propagates throughout the underlying nematic causing a weak macroscopic twist \cite{lubensky1998}.
The result is a chiral nematic phase, also called \textit{cholesteric},
possessing \text{mesoscopic} chirality characterized by a mesoscopic \textit{pitch},
\textit{i.e.} the distance at which the nematic director revolves by a full turn, and handedness indicating the direction of twist (left or right).

One of the most emblematic chiral shapes is the helix.
Prominent examples can be found on the macromolecular scale  such as the double-helix structure of DNA  \cite{kornyshev2007a} and the alpha-helix secondary structure of proteins.
Helical shapes may also arise due to spontaneous assembly of smaller chiral molecular units such as in the case of cellulose nanocrystals (CNCs) which adopt the shape of a twisted nanoribbon \cite{gray-cullulose,kelly2014}.  These materials offer a promising route towards the
fabrication of security papers or tunable mirrorless lasers \cite{lagerwall2014a,schutz2015}. Helices are also abundant in the biological world; microtubules possess a helical internal structure \cite{hunyadi}, and the rigid spiral shape of bacterial flagella can be exploited to study the effects of shape chirality on the structure and dynamics of dense fluid phases {\cite{barrybeller}.

While the helicity of macromolecules is generally fixed by the molecular geometry of the atomic and molecular bonds,
the origin of supramolecular helicity is often of a liquid crystalline nature \cite{gennes-prost}. The mechanism underpinning  the geometrical properties of the helical director field and its handedness are, however, rather unclear even in simple systems and one may ask:
``How does the helicity propagate from the macromolecular to the supramolecular scale?" and
``What fundamental physical quantities control their left- or right-handed symmetry? 

\footnotetext{\textit{$^{a}$~
Laboratoire de Physique des Solides, CNRS, Univ.\ Paris-Sud, Universit\'{e} Paris-Saclay, UMR 8502 -- 91405 Orsay Cedex, France
E-mail: rik.wensink@u-psud.fr}}

The nature of the helical twist of the director field can be experimentally controlled
by the properties of the macromolecules as well as by certain thermodynamic properties such as temperature, particle density, or solvent conditions \cite{katsonis2012a}.
There is a large body of experiments on cholesteric systems composed of chiral biomolecules revealing a particularly sensitive relationship between supramolecular helical order and external control parameters.  For example, chinin suspensions extracted from crab exoskeletons
form a cholesteric whose pitch is controlled
by a subtle balance of the packing fraction and ionic strength \cite{belamie2004}.
Microtubules in the cytoskeleton \cite{kornyshev2007a}
are self-assembled from a collagen triple helix fibers \cite{giraud2008a}.
Concentrated solutions of DNA \cite{livolant1996, strey-dna}
form both left- and right-handed cholesterics with different pitch versus temperature trends depending
on the sequence of oligomers \cite{zanchetta2010a}.
Spontaneous helix inversions upon change of temperature were observed in thermotropic systems of polypeptides \cite{watanabe_polypep} and cellulose \cite{yamagishi},
and inversions upon changing the solvent conditions were reported for lyotropic systems poly-benzyl-L-glutamate \cite{toriumi} indicating that the nematic twist is governed by a subtle balance of forces whose exact origin remains partly obscure.

In this paper, we use large scale molecular dynamics simulations of a simple model system consisting of an achiral  spherocylindrical backbone dressed with soft Yukawa patches forming a helical arrangement. Our particular focus is on finding evidence for so-called \textit {pitch inversions}, where a sudden change of the cholesteric sense is produced upon
a change of particle density at fixed molecular pitch. These inversions have recently been conjectured from Onsager-type theories for chiral nematics of both soft \cite{wensink2014} and hard helices \cite{dussi2015a} suggesting that these inversions are somehow inherent to a nematic organization of helical nanoparticles. These theories are, however, based on the second-virial approximation and do not  capture particle correlations beyond mere pairs \cite{Onsager}. Moreover they are {\em mean-field} theories in which thermal fluctuations of the particle density, orientations and fluctuations in the (local) nematic director orientation are neglected.  A careful molecular simulation study capable of accounting for these effects is therefore necessary and timely, in particular, in view of the subtle balance of torques that underpin these helical inversions.

Our inspiration for focussing on long-ranged chiral interactions stems from the observation of chiral nematic order in
colloidal suspension of filamentous {\em fd} virus rods \cite{dogic2000a}.
\citet{grelet2003a} have attempted to study the mechanism
behind the twist by coating the {\em fd} virus with a thick layer of polymer in order to suppress the
short-range chiral interactions and to treat the virus as a sterically stabilized colloid.
The nematic phase, however,  remains twisted and the pitch depends markedly on the salt concentration
suggesting that the supramolecular twist must be due to weak but long-ranged electrostatic forces reaching beyond the distance shielded off by the polymer coating.
A related study on another filamentous virus (M13)  \cite{tombolato2006a} argues that the observation of
left-handed cholesteric phases formed by right-handed chiral viruses
is also due to a subtle competition between electrostatic interparticle forces. A similar conclusion is reached in the case of chiral phases of DNA \cite{tombolato2005a,frezza2011a}.

Although the focus of this paper is on the competition between short-range steric and long-range chiral electrostatic repulsions
it is worth mentioning the extensive body of work done on hard helices \cite{frezza2013a,kolli2014a,kolli2014b,kolli2016a},
where chirality is transmitted  via steric interactions emerging  from the helical shape \cite{frezza2014a,belli2014a,dussi2015a,wensink2015a}.
Following up on the early work of \citet{straley1976} on hard threaded rods,
 density functional theory (DFT) was used to show
that right-handed helices yield left-handed cholesterics at large microscopic pitch,
and right-handed cholesterics if the microscopic pitch is small \cite{frezza2014a,dussi2015a}.
This behavior is ascribed to small excluded-volumes differences between left- and right-handed twist on the pair level.
Whilst the geometry of the hard helix is the main parameter defining the sense of orientation,
the degree of local alignment (as determined by the particle density) also plays a crucial role and may bring about pitch inversions at fixed internal helicity  \cite{belli2014a}. 
We reproduce these pitch inversions in our model for soft chirality and confirm their robustness with respect to fluctuations and multi-particle correlations that were hitherto neglected in these theoretical studies.

\begin{figure}
   \centering
  \includegraphics[width= 0.8 \columnwidth]{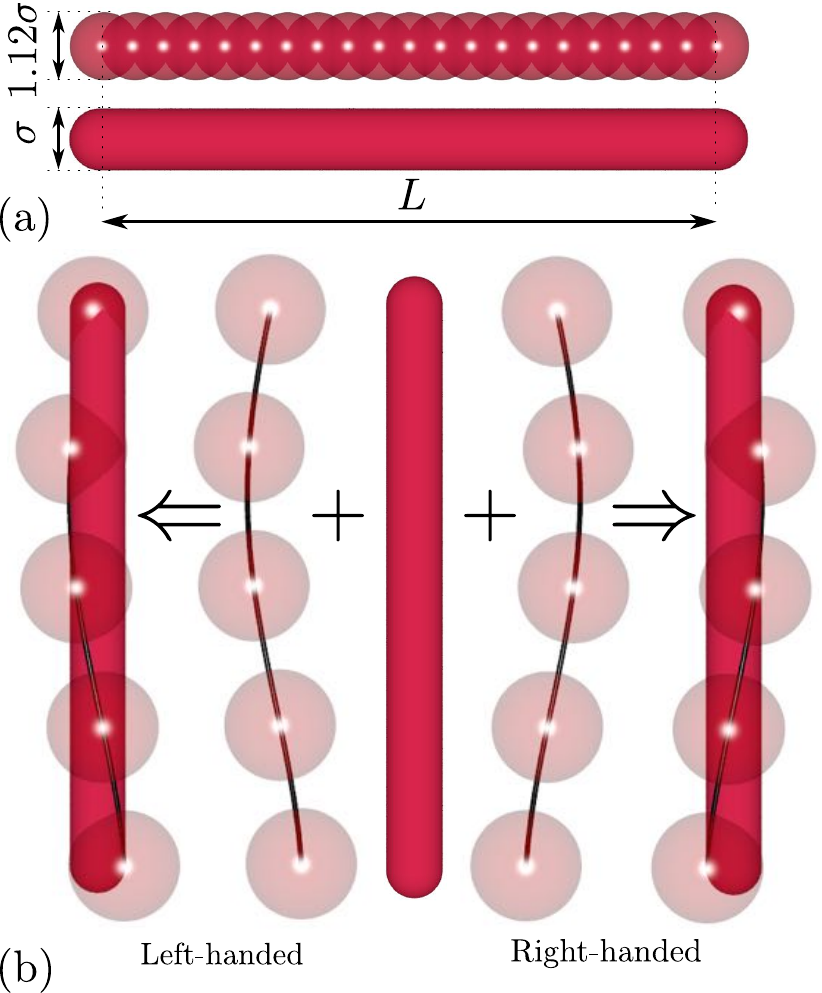}
  \caption{Geometry of a helical patchy rod.
  (a) Each rod is composed of 21 fused WCA spheres representing a spherocylinder with aspect ratio $L/\sigma = 10$.
  (b) A particle is composed of a bare achiral backbone supplemented with a left- or right-handed helix.
      The helix is discretized into a
      finite number of repulsive Yukawa patches represented as soft spheres.
      The interaction range of the patches equals twice the sphere diameter.}
  \label{fig.geomic}    
\end{figure}

Particle-based simulation of chiral nematics is a non-trivial task and has been a long-standing challenge in computer modelling of liquid crystals \citep{allen1993,frenkel2013a}.
The first complication is that the cholesteric pitch  is usually much larger than the size of the simulation box. This requires simulating very large systems even though the computational burden can be alleviated somewhat using twisted boundary conditions, or by using a simulation box taking the shape of a hexagonal prism which accommodates for larger pitches \cite{allen1993}. The second, more serious problem is that the two opposing planes perpendicular to the helix axis
are periodic. This implies that, unless the system dimension along the helix axis exactly matches an integer times half the cholesteric pitch, the periodicity of the boundaries is incompatible with the natural helical order and a spurious torque is imparted onto the system.
Two main workarounds have  been developed to allow the pitch to be measured without the aforementioned bias.
The first is replacing the two sides perpendicular to the twist direction by two opposing hard walls, thus confining the system to a slab geometry. This breaks the periodicity along the helix axis and lifts the undesired interactions introduced by the periodic boundary conditions \cite{brumby2010a,varga2006a,varga2003a}. A disadvantage of this method is that the system loses its bulk properties near the boundaries due to the depletion forces exerted by the walls. 
The second approach, by contrast, takes advantage of the unphysical strain imparted by the periodicity
and samples the average torque density the system experiences when the rods are constrained to adopt an unwisted (nematic) or a twisted configuration (nematic director making a half turn across the box dimension) \cite{germano2002a,allen2001a}. Assuming the equilibrium twist to reside somewhere between the imposed untwisted and overtwisted one, one can compute the equilibrium pitch (and also the twist elastic constant)  simultaneously via a simple interpolation procedure. 

Most simulation studies of cholesterics reported so far rely on pairwise interactions
which can be split into chiral and achiral parts.  The intermolecular potential is usually based on some Gay-Berne \cite{memmer1993,germano2002a} or hard-spherocylinder \cite{varga2003a,varga2006a,brumby2010a} reference potential
supplemented with a simple pseudoscalar-type interaction potential encoding chirality. Alternatively, chirality can be introduced by the addition of chiral dopants \cite{earl2004a} or induced by a chiral surface \cite{berardi_zannoni1998}.
The routinely used pseudoscalar potential is similar to the one
proposed for soft helices \cite{wensink2014}  with the important difference that, in the latter, the molecular twist and handedness follow directly from the geometric features and interaction potential of the helices rather than having to be {\em prescribed} as  input parameters.
To our knowledge, the present paper provides the first simulation study aimed at
measuring helical pitches for a simple model of chiral patchy cylinders 
where the interactions can be split into a achiral and achiral part,
and where the molecular pitch can be controlled explicitly
by subtle changes in the spatial arrangement of the patches.

\section{Simulation model}

In our model, each molecule is composed of a hard rodlike backbone superimposed with a Yukawa helix \cite{wensink_jpcm2011,wensink2014},
as illustrated in Fig.~\ref{fig.geomic}.
The rod is modelled as a linear rigid body
composed of 21 fused spheres, each interacting with the spheres from neighbouring rods via a Weeks-Chandler-Andersen potential
\begin{equation*}
V_\text{WCA}(r) =
\begin{cases}
4\epsilon\left[ \left( \sigma/r\right)^{12} - \left(\sigma/r\right)^{6} \right] + \epsilon & r < 2^{1/6}\sigma, \\
0 & r \ge 2^{1/6}\sigma,
\end{cases}
\end{equation*}
where $r$ is the centre-of-mass distance between the spheres and $\sigma$ is the approximate diameter of the repulsive core
and $\epsilon$ is the interaction strength  in units of the thermal energy $k_{B}T$ (in terms of Boltzmann's constant $k_{B}$ and temperature $T$).
We take $\sigma = 1.0$, thus serving as our unit of length, and fix $\epsilon =1.0$.
The rod length is set at $L = 10\sigma$.
The helix is discretized into $n_p$ repulsive Yukawa patches with the potential
\[ V_\text{Y}(r) = A_p e^{-\kappa r}/r, \]
where $\kappa = 2.0\sigma$ is the screening length.
The potential is cut-off and shifted to zero at $r_c = 5\sigma$.
To ascertain that the total Yukawa repulsion between two molecules is constant
and independent of the discretization of the spiral,
the patch amplitude $A_p$ is normalized by the number of patches via $A_p = A/n_p^2$.
Unless otherwise stated, $n_p = 5$ for all systems simulated.
The centres of the patches are located on the surface of the rod (distance $\sigma/2$ from the main axis)
and its helical layout is described by a microscopic pitch $p$ which can be right-handed or left-handed.
For reasons of basic symmetry we only consider rods with a left-handed microscopic pitch.

\begin{figure}[htb]
\centering
  \includegraphics[width= \columnwidth]{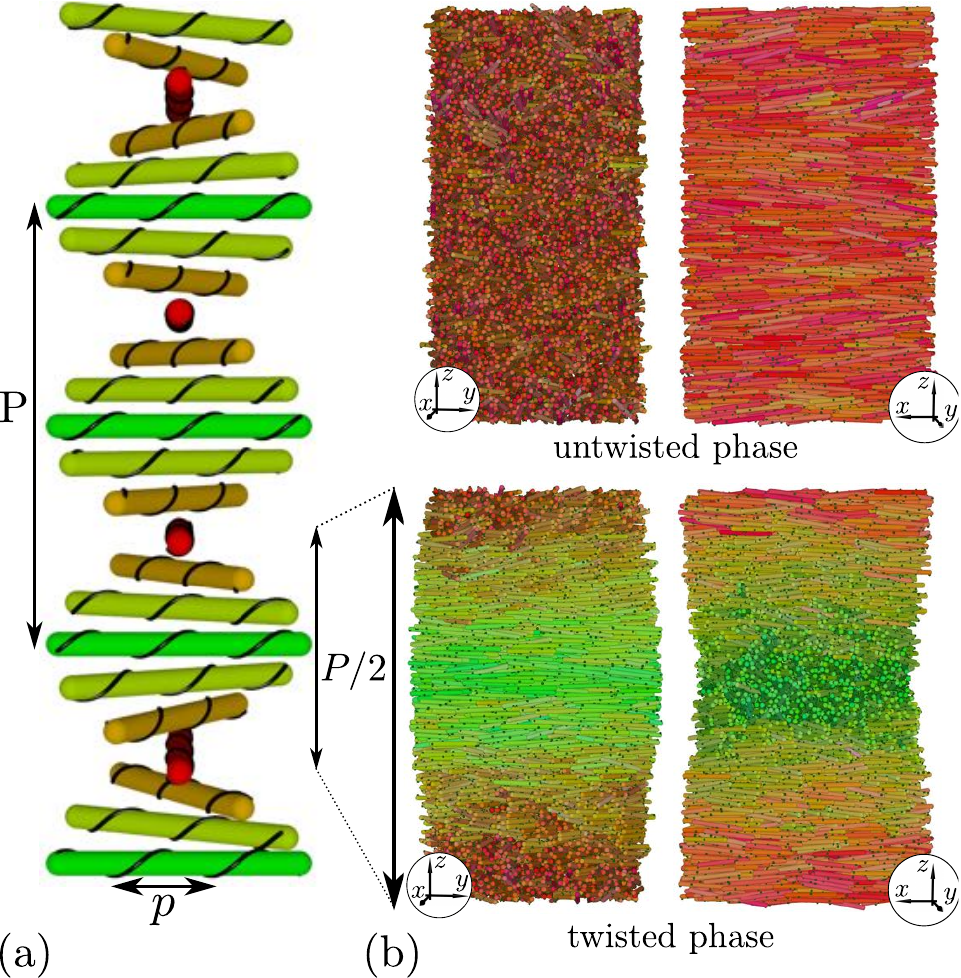}
  \caption{Snapshot of the untwisted and twisted systems where the patchy rods are colored according to their orientations.
  (a) A schematic illustration of the different length scales of the system; the microscopic pitch with magnitude $p$ and the supramolecular, cholesteric pitch with length $P$.  In this representation the main rod direction is identified with the local nematic director.    
      The molecular pitches considered in this paper are of the scale of rod length $p \sim L$, whereas the macroscopic pitches are much larger than the rod length ($P\gg L$).
  (b) Snapshots of the untwisted and twisted states viewed from two different angles.
      The director in the twisted phase revolves by a half turn along the $z$-direction. The boundaries along the $z$-axis are fully periodic.
      The centers of Yukawa patches are indicated by
      small dark dots on the surface of the rods. The pitch of the twisted state has magnitude $2L_z$ and is significantly smaller
      than the equilibrium pitch $P$, which typically ranges from $10L_z$ to $50L_z$. }
  \label{fig.geomac}    
\end{figure}

To sample the statistics of the many-body system we use a dynamical integrator for rigid bodies \cite{kamberaj2005a,miller2002a},
which has recently been implemented within the molecular simulator LAMMPS \cite{plimpton1995a}.
The system is composed of $N=6534$ rigid molecules in a canonical ($NVT$) ensemble
and periodic boundary conditions in all three Cartesian directions are applied. The simulation box is rectangular with dimensions $L_z=2L_x=2L_y$
and volume $V = L_xL_yL_z$.
We define the \textit{packing fraction} of the bare rods via
$\phi = N(\pi \sigma^3/6 + \pi L \sigma^2/4)/V$
ignoring any contributions from the patches
and assuming that the segmented rod can be mapped onto a hard spherocylinder \cite{mcgrother1996a}.
For simplicity the terms \textit{density} and packing fraction both refer to $\phi$.
The time step in all simulations is fixed at $\Delta t =  0.003$.

As alluded to in the introduction, simulating a twisted nematic without bias from the periodic boundaries
requires the macroscopic half pitch length
to be exactly a multiple of the box dimension \citep{allen1993}.
Since we do not know the equilibrium pitch \textit{a priori} an alternative strategy must be implemented. Here, we follow the procedure of \citet{allen2001a} which consists of measuring the average torque density of an ensemble of rods adopting an artificial twist imposed by the periodic boundaries. 
As shown in Fig.~\ref{fig.geomac},
the system is simulated in two metastable states
with macroscopic twist defined by a wavenumber
$k = 0$ (untwisted, nematic) and $k = -\pi/L_z$ (half turn twist),
with the latter corresponding to the left-handed macroscopic twist.
Although compatibility with the periodic boundary condition along the $z$-direction is guaranteed in both cases, the systems are both {\em metastable}  \citep{germano2002a,allen2001a}
because the imposed twist will be different from the equilibrium one.

The equilibrium pitch can be computed as follows. Let us  denote the $z$-component of the torque vector
between particle $i$  and $j$ as $\tau_{ij}^z$.
Similarly, we denote $r_{ij}^z$ as the $z$-component of the distance vector
between the centre of particle $i$ and $j$.
From this we define the tensor component
\[ \Pi_{zz} = - \frac{1}{2} \sum_{i = 1}^N \sum_{j \neq i }^{N} r_{ij}^z \left( \tau_{ij}^z - \tau_{ji}^z \right), \]
representing the torque per unit area.
We denote the averages of $\Pi_{zz}$ as $\langle \Pi_{zz} \rangle_0$
for the untwisted sample and $\langle \Pi_{zz} \rangle_k$ for the half-turn twisted one with $k = - \pi/L_z$.
The equilibrium macroscopic pitch can then be obtained from
\begin{equation}
P = -\frac{2\pi}{k}
     \frac{\langle \Pi_{zz} \rangle_k - \langle \Pi_{zz} \rangle_0}{\langle \Pi_{zz} \rangle_0 }
  = -\frac{2\pi V}{\langle \Pi_{zz} \rangle_0} K_{2},
\label{eq.p}
\end{equation}
The pitch $P$ is right-handed if $P>0$ and left-handed if $P<0$.
An important advantage of the method is that it allows for a simultaneous measurement of the twist elastic constant $K_{2}$ which follows from
\begin{equation}
K_2 = \frac{\langle \Pi_{zz} \rangle_k - \langle \Pi_{zz} \rangle_0}{Vk}.
\label{eq.k2}
\end{equation}
It turns out that, with the setup outlined in Fig.~\ref{fig.geomac}, the tensor components $\Pi_{zz}$ can be measured without difficulty since the metastability of the untwisted and overtwisted states is sufficiently strong to guarantee good statistics.

All simulations are initiated from a cubic crystal composed  of molecules
with random azimuthal angles and with their
main axes aligned along the $x$ direction.
The untwisted nematic is generated by melting the crystal
during an equilibration run of $2\cdot10^6$ time steps.
The twisted state is prepared
by imposing two hard walls at the top and bottom side of the simulation box
 and by applying an external torque onto the rods, consisting of two force fields of opposite sign (force dipole),
$-\vec{F}(z)$ and $\vec{F}(z)$, acting on either end segment of the spherocylinder.
The direction of the force describes a half turn along the $z$ axis and is
parameterized by  $\vec{F}(z) = 100\cdot\left(\cos(\pi z/L),\sin(\pi z/L), 0 \right)$ with $0 < z < L_z$.  A cubic crystalline starting configuration is then allowed to relax during an equilibration run of $10^6$ steps
while the imposed external fields enforce the system to adopt the desired half-turn twist with the local director residing in the $xy$-plane of the simulation box.
The external torque field and the confining walls are subsequently removed,
and the system is equilibrated for another  $2\cdot10^6$ steps while remaining in the twisted state.
The length of the production run is $5 \cdot 10^6$ steps
with averages of $\Pi_{zz}$ calculated every $5\cdot10^3$ steps.
The simulation runs at a speed of $10^6$ steps which takes about 15 hours on 4 processors.
The chosen system size is sufficiently large to ensure that finite size effects are negligible in our analysis.

\section{Results}
 
\subsection{The choice of parameters and phase behavior}

Since our focus is on nematics, we must restrict the range of packing fractions to the  relevant region where nematic order is stable against smectic order at large packing fraction, and against isotropic order at low density.
An additional requirement is that the density must be high enough to keep the constrained twisted nematic stable during the production run. We find that this stability is no longer guaranteed at densities close to the nematic-isotropic transition.  In what follows, we briefly discuss the choice of parameters,
in particular the rod geometry and its effect on the phase behavior.

If the rod  aspect ratio is small
the density region over which the nematic is stable is narrow \cite{mcgrother1996a},
and the metastability of the twisted nematic is too weak.
If the rods are too long, the simulations become computationally expensive
because larg systems with large nearest neighbor cells need to be considered.
We find that  $L = 10\sigma$ provides the optimal aspect ratio.
In view of the conditions above, we consider packing fractions in the range $0.34 \le \phi < 0.43$.
The upper bound of this interval is the nematic-smectic transition density
of the achiral patchless rods, which is in a very good agreement with earlier predictions for hard spherocylinders \citep{mcgrother1996a}.
The lower bound is a safe estimate
of the lowest density at which the imposed twisted nematic can be kept stable during the course of the simulation run.

As for the phase stability of the nematic, the addition of repulsive patches to the spherocylindrical backbone
produces two distinct effects. 
First, the effective aspect ratio is reduced.  The phase diagram of hard spherocylinders \cite{bolhuis1997,mcgrother1996a,wittmann2015a} then implies
that the interval of the nematic stability should become narrower
and shift to higher packing fractions.
Second, the patches  increase the effective density of the system.
We find that the increased density
has a more pronounced impact on the phase boundaries
than the reduction of the aspect ratio.
The net implication is that both the isotropic-nematic and nematic-smectic transition densities
shift to smaller packing fractions $\phi$ when the amplitude $A$ grows larger.
We found that the phase boundaries are also affected by changes in the microscopic pitch.
For example, a nematic system with $A = 1000$, $n_p = 5$ and $\phi = 0.37$
crosses over into a smectic for large microscopic pitches $p > 15$.
We stress that all results shown in this paper are in the stable nematic range. Data points corresponding to systems that are believed to be affected by long-lived smectic fluctuations are indicated by open symbols in all figures.

We finally mention that  kinetically arrested structures such as a Wigner glass \cite{meijer1991,sciortino2004} may occur in systems with long range interactions.
To ensure that we steer clear of dynamically arrested states,
we monitor the diffusive properties of the particles
at the densest phases simulated.
The translational diffusion constant is defined as $D = \lim_{t \rightarrow \infty} \langle r^2(t)\rangle/6t$,
where $\langle r^2(t)\rangle$ is the mean square displacement.
The smallest diffusion constant we measure is $D = 5\cdot10^{-5}$ for a system
with $A_p = 20$ and $\phi = 0.42$. This implies that during the production run, a rod traverses an average distance of $25D = 2.5L$ 
and that there are no signs of dynamic arrest.

\subsection{Pitch versus density: spontaneous sense inversion}

Let us first address the sensitivity of the pitch on the  discretization of the Yukawa spiral.
The results in Fig.~\ref{fig.dis} show that the use of 5 patches constitutes a good approximation
of a continuous spiral  \cite{wensink2014}, in which case the number of patches would be infinite.
We find that using a larger number of patches puts a considerable burden on the simulation
time, due to the long-range interactions, with only limited extra accuracy.

\begin{figure}
   \centering
   \includegraphics[width= 0.8 \columnwidth]{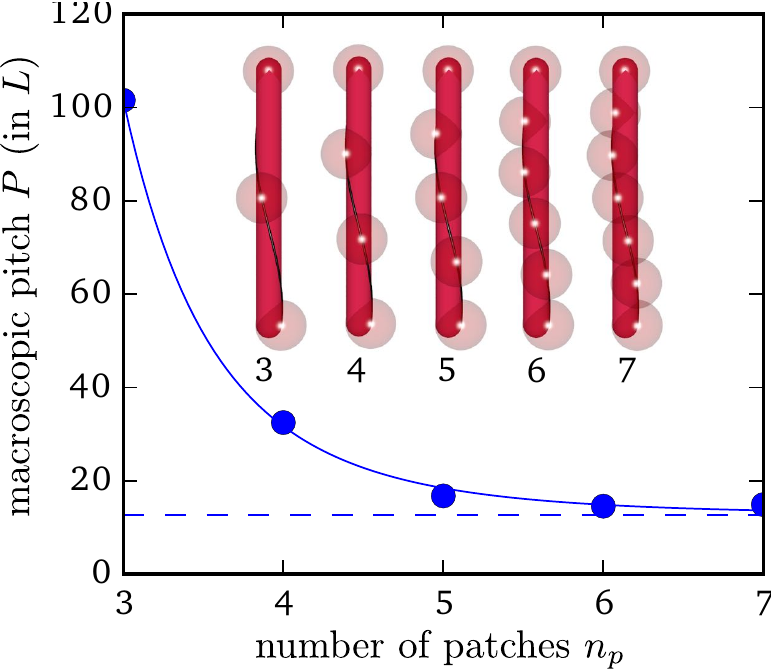}
   \caption{Dependence of the macroscopic pitch on the discretization of the Yukawa spiral  into  $n_p$ Yukawa patches. The two-patch rod model ($n_p = 2$) is non-chiral and produces a large pitch $843L$ resembling an untwisted nematic.
             All systems have a packing fraction $\phi = 0.35$, amplitude $A = 1000$, and microscopic pitch $p = 1.25L$.  The solid line is drawn to guide the eye.}
   \label{fig.dis}
\end{figure}

\begin{figure}[htb]
\centering
  \includegraphics[width= 0.8 \columnwidth]{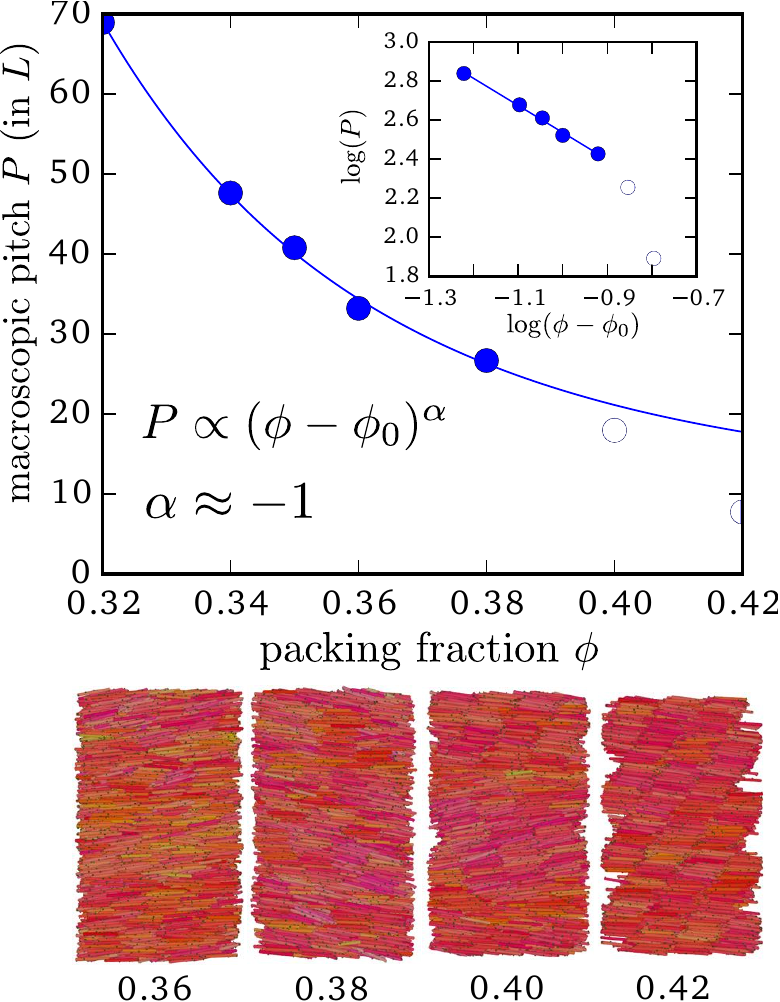}
  \caption{
 Variation of the macroscopic pitch with packing fraction for
  a system with $A_p = 20$ and $p = 1.2L$.
  The fit is shown on a double log-scale (inset) with the isotropic-nematic transition density estimated as $\phi_0 \approx 0.26$.
  Shown below are  snapshots  at different densities
  close to the nematic-smectic transition.}
  \label{3441Pphiall}
\end{figure}

\begin{figure}[htb]
\centering
  \includegraphics[width= 0.8 \columnwidth]{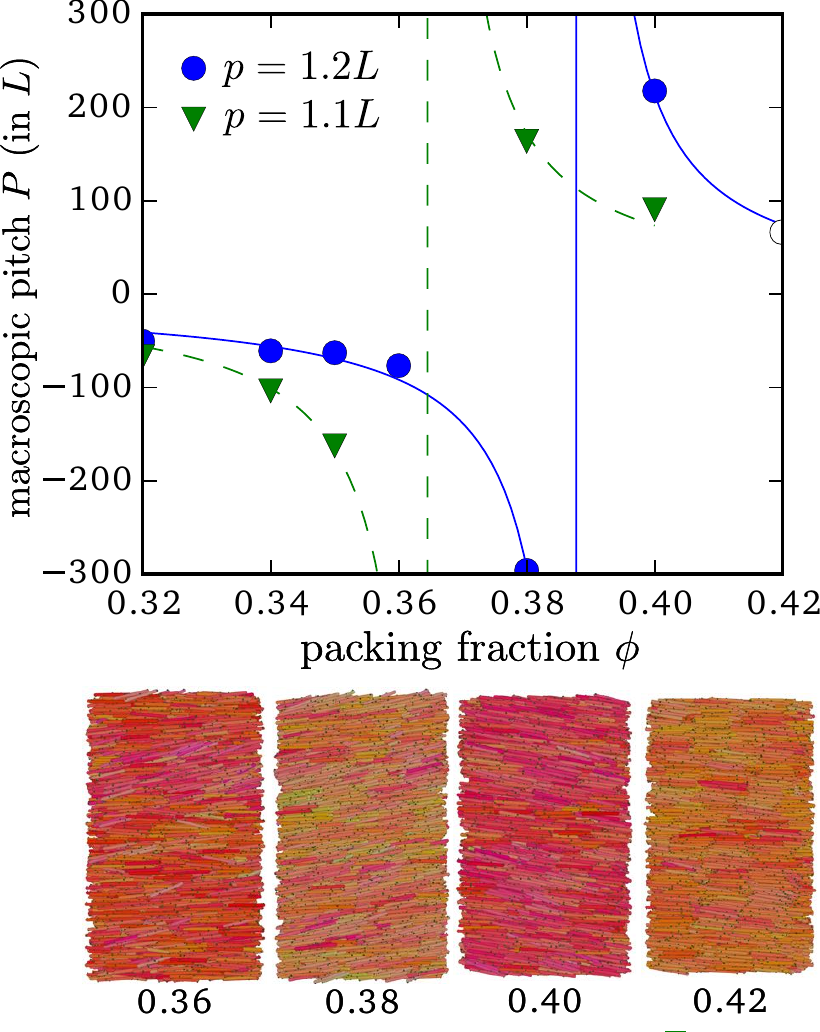}
  \caption{
  Cholesteric pitch inversion as a function of the packing fraction
  for chiral amplitude $A_p = 10$ and  two different values of microscopic pitch $p$.  
  Below are the typical snapshots of the system at different densities and $p = 1.2L$.
  The last snapshot exhibits a slow onset of smectic fluctuations. Drawn lines are curves fitted on the form $P = P_{0}(\phi - \phi^{\ast})^{-1}$ with offset $P_{0}$ and inversion density $\phi^{\ast}$. }
  \label{3433Pphiall}
\end{figure}

Next, we scrutinize the effect of particle density.  The results in Fig.~\ref{3441Pphiall} exhibit a common trend where the cholesteric pitch decreases systematically with concentration.
The scaling  of the pitch with density around the isotropic-nematic transition density $\phi_0$ reveals  a simple power law
\begin{equation}
P \propto (\phi - \phi_0)^\alpha,
\label{eq.ppa}
\end{equation}
with exponent $\alpha \approx -1$ in quantitative accord with theoretical predictions \cite{odijkchiral,wensinkjackson}
and  in qualitative agreement with experimental results on {\em fd} virus suspensions at reduced salt concentration \cite{grelet2003a} .

Examples of spontaneous inversions of the macroscopic helical symmetry can be gleaned from Fig.~\ref{3433Pphiall},
Here, the direction of twist suddenly changes from left to right upon variation of the density.
Comparing the low density regions of Figs.~\ref{3433Pphiall} and \ref{3441Pphiall}  reveals that the twist is systematically negative for the weaker amplitude,
and positive for the stronger amplitude. The effect of the amplitude on the helical symmetry is further examined in Fig.~\ref{fig.varinvs2}(a).
It is very plausible that the helical inversion with amplitude originates from packing effects as well (cf.  Fig.~\ref{3433Pphiall})
given the enhanced effective system density at increasing amplitude.

The density at which the inversion occurs can be lowered by a slight reduction of the microscopic pitch as is evident from Fig.~\ref{3433Pphiall}.
This observation tallies with predictions from DFT \cite{wensink2014} where the critical degree of local alignment (proportional to rod density) at which the inversion takes place drops upon decreasing the molecular pitch. A remarkable feature is that the magnitude of the macroscopic pitch $P$ should strictly diverge at the inversion point where the cholesteric should unwind completely and form a nematic phase even though the rods are distinctly chiral.
None of the systems with a density close to the inversion point are affected by fluctuations pertaining to smectic layering as can
be seen from the snapshots attached. There is no sign of an anomalous increase of the twist elastic resistance near the critical density as will be demonstrated in the next paragraph.

The handedness of the macroscopic pitch $P$ can also be changed by modifying the microscopic pitch $p$. The direct correlation between the microscopic and macroscopic pitch is highlighted in Fig.~\ref{fig.varinvs2}(b) and confirms the interrelationship to be highly non-trivial. More importantly, it illustrates the possibility of cholesteric helix inversions induced by changes in the molecular conformation affecting the internal helicity of the constituents.  These conformational  modifications may, in turn, be governed by changes in temperature in case of thermotropic assemblies.

\begin{figure}[htb]
\centering
  \includegraphics[width= 1\columnwidth]{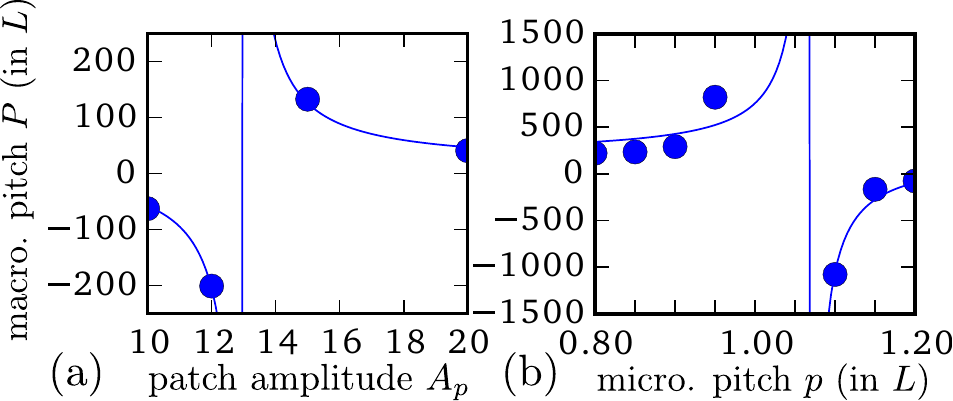}
  \caption{
  (a) Pitch inversion as a function of the strength of Yukawa amplitude $A_{p}$ 
      for $\phi= 0.35$ and $p = 1.2L$.
  (b) Pitch inversion with respect to the magnitude of the microscopic pitch 
      for $\phi = 0.36$ and $A_p = 10$.}
  \label{fig.varinvs2}
\end{figure}

Some insight into the microscopic origin of the pitch inversion can be gathered from analysing
the torque per unit area in the untwisted phase.
Equation~\eqref{eq.p} implies that an inversion ($|P| \rightarrow \infty$) occurs when the net twist propensity vanishes, i.e., $\langle\Pi_{zz}\rangle_0 \rightarrow 0$.
Assuming a pairwise additive interaction, we can split the torque into a chiral part corresponding to the Yukawa patches and a nonchiral part representing the backbone segments, 
$\langle\Pi_{zz}\rangle_0 = \langle\Pi_{zz}^\text{a}\rangle_0 + \langle\Pi_{zz}^\text{c}\rangle_0$. A pitch inversion can then be identified with a balance of mean torque densities: $\langle\Pi_{zz}^\text{a}\rangle_0 = - \langle\Pi_{zz}^\text{c}\rangle_0$.
The results in Figures~\ref{fig.varinvs}(a) and (b) show that $\langle\Pi_{zz}^\text{a}\rangle_0$ remains negative throughout the density interval while $\langle\Pi_{zz}^\text{c}\rangle_0$,
changes sign going from small to large density.
The rapid upswing from negative to positive observed in the chiral contribution $\langle\Pi_{zz}^\text{c}\rangle_0$
outweighs the negative contribution of the achiral torque $\langle\Pi_{zz}^\text{a}\rangle_0$  and causes the cholesteric to twist  from left- to right-handed. From this we can conclude that  the sense inversion is entirely driven by the chiral part of rod potential. Moreover, the intermolecular torque generated by the helical patches may either have a positive or negative signature depending on the density of the underlying nematic structure as put forward by theory \cite{wensink2014}.

\begin{figure}[htb]
\centering
  \includegraphics[width= 0.9\columnwidth]{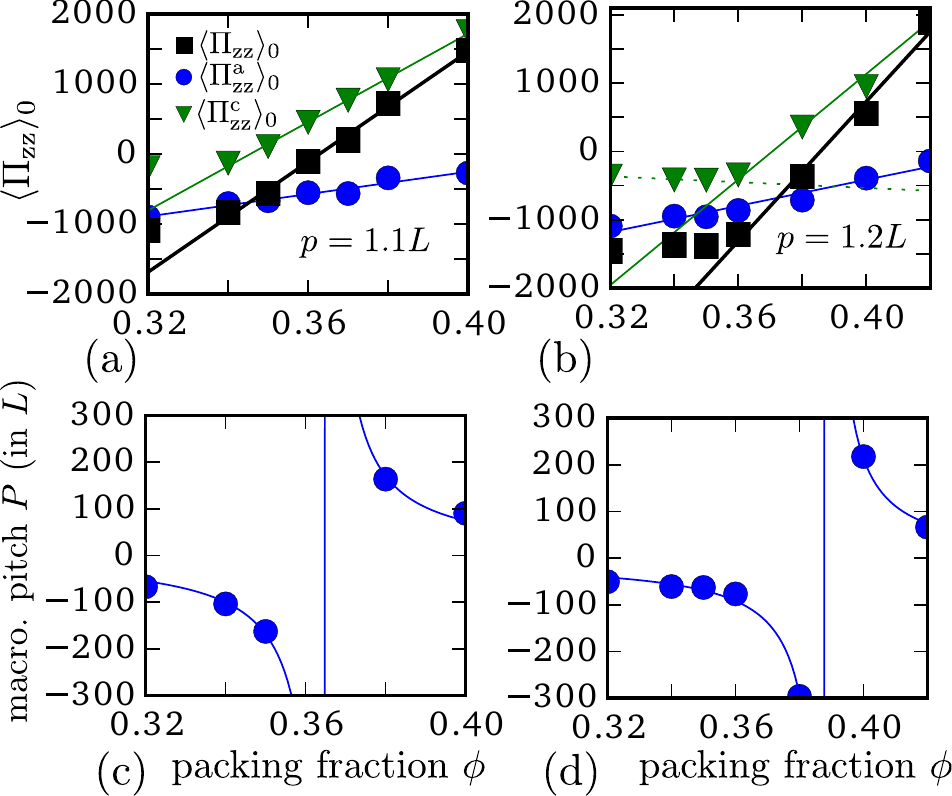}
  \caption{(a,b) Torques versus packing fraction for untwisted systems with $A_p = 10$
                    and two different values of the microscopic pitch $p$. 
           (c,d) Corresponding macroscopic pitches $P$ (cf. Fig.~\ref{3433Pphiall}).}
  \label{fig.varinvs}
\end{figure}

Close inspection of Fig.~\ref{fig.varinvs}(b) reveals that $\langle\Pi_{zz}^\text{c}\rangle_0$ has a local minimum
just before the onset of the fast growth,
where the linear fit yields a negative gradient at low densities
and positive gradient at high densities.
This suggests the existence of an optimum density in the transition region
where the negative twist is maximal and hints at a sudden emergence of a positive torque contribution at a critical density.

In case of a scenario with no pitch inversion (Fig.~\ref{3441Pphiall}),
the contribution of the chiral part $\langle\Pi_{zz}^\text{c}\rangle_0$
is strictly positive (results are not shown) and their values are significantly larger
than the contributions of the achiral part $\langle\Pi_{zz}^\text{a}\rangle_0$,
which are systematically negative. Therefore, in this situation the orientation of the macroscopic pitch remains right-handed throughout
the nematic density range.

\begin{figure}[htb]
\centering
  \includegraphics[width=0.9\columnwidth]{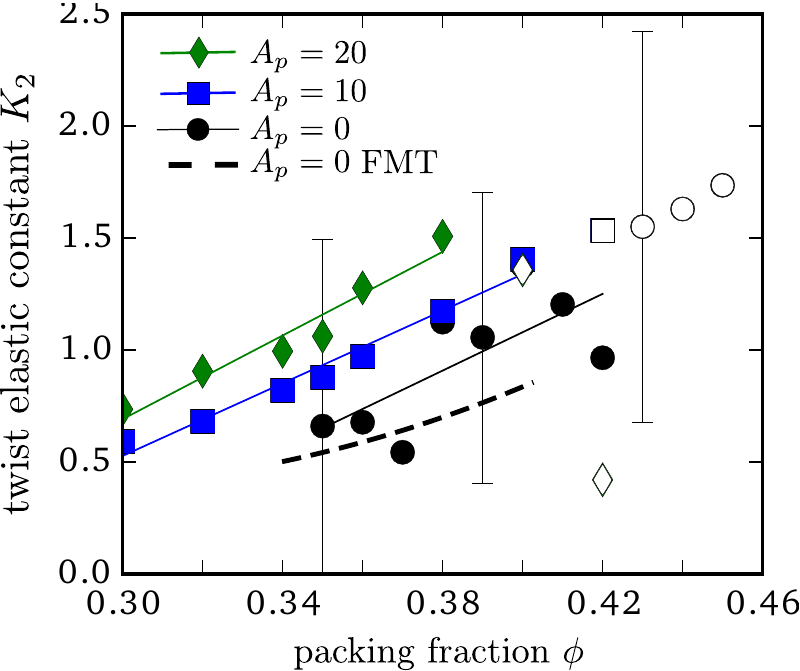}
  \caption{
  Twist elastic constant versus packing fraction for rods with microscopic pitch $p = 1.2L$,
  and different values of the patch amplitude $A_p$.
  Filled symbols represent simulation results. The empty ones correspond to smectic samples.
  The values for $A_p = 0$ (patchless rods) are in fair agreement with the prediction from fundamental measure theory (FMT) for hard spherocylinders \citep{wittmann2015b}.
  Error bars are calculated from the covariance of $10^3$
  measurements of $\langle\Pi_{zz}\rangle_k$ and $\langle\Pi_{zz}\rangle_0$. 
  The error bars for $A _{p}> 0$ are not displayed but are of comparable magnitude.
  }
  \label{fig.k2}
\end{figure}

\subsection{Twist elastic constant versus packing fraction}

It is worthwhile to briefly embark on a detailed analysis of the twist elastic constants $K_{2}$ emerging from our simulation data. This is particularly so in view of the complications encountered  in measuring  the elastic properties of liquid crystals in simulation \cite{allen1993,allen1993b,humpert2015a} and because of their importance in the interpretation of experimental results \cite{gennes-prost}. In fact,  as far as we are aware, no simulation results have been published as yet showing how the twist elasticity of spherocylinder nematics behaves with density. 
Similar to the pitch, the elastic constant can be readily obtained from the torque density contributions according to Eq.~\eqref{eq.k2}.
The results for the bare spherocylinders are shown in Fig.~\ref{fig.k2}
and compared with predictions from fundamental measure density functional theory \cite{wittmann2015b}.
It is striking to observe that $K_{2}$ seems to grow linearly with density without showing any sign of divergence close to the nematic-smectic transition as has been often conjectured in view of the generic incompatibility between lamellar order and twist \cite{degennes-elastic, pindak}.
Our results, however, suggest that the slow growth with  density
extends smoothly to regions where  smectic fluctuations become predominant.
A similar trend is observed for  the patchy rods; although the amplitude of $K_{2}$ is enhanced upon increasing the amplitude of the Yukawa patches, its (near-)linear density dependence is preserved. We stress that in view of the large errors incurred and because of finite size effects that  may play a role at larger packing fractions where smectic fluctuations prevail the results in Fig.~\ref{fig.k2} can only provide a qualitative, yet valuable guideline.  

\subsection{Pitch measurement in slab geometry}

In addition to the torque-based technique
we have also attempted to determine the macroscopic pitch
using  Monte Carlo simulations of a similar model (hard spherocylinders plus helical Yukawa patches)
in which the rods are confined between two hard walls \cite{varga2003a, brumby2010a}.
The walls enforce the nematic to align along the $xy$-plane
and at the same time break the periodicity along the $z$-axis,  thus ensuring that particle orientations are not coupled through the boundaries imposed along the twist direction.
The macroscopic pitch can then be estimated from the twist of the local nematic director $\vec{n}(z)$ by sampling bulk statistics far from the boundaries.
However, in our simulations, we observe large fluctuations in the density close to the walls,
which point to the formation of a smectic wetting layer.
These smectic layers show signs of dynamic arrest and 
misalignment across the periodic boundaries in the $xy$-plane. As a consequence, they impart a strong spurious twist onto the system and suppress the natural weak twist favored by the Yukawa spirals.
Even though the effect of the hard walls
is negligible around the center of the simulation box, where bulk behavior should be recovered,
we believe that the presence of these long-lived smectic fluctuations
undermines the suitability of the hard-wall method for measuring
macroscopic pitches in a bias-free manner, at least for the model system and concentration ranges investigated here.

\section{Conclusions}

We have presented large-scale simulations of cholesteric phases of helical patchy rods  and provide evidence 
of a delicate relation between  chirality on the molecular scale and the macroscopic scale.  In particular, we have established that the magnitude and handedness of the cholesteric pitch can be carefully tuned by the packing fraction of the system, by the magnitude and sign of the microscopic pitch and that it shows a marked sensitivity with respect to the strength of the chiral interactions.
While left-handed helical rods twist into left-handed cholesterics at low packing fractions (or at weak chiral interactions)
they form right-handed cholesterics at higher packing fractions (or at strong chiral interactions).

We show that the density at which the pitch inversion occurs depends the microscopic pitch; the observed trend is in line with recent theoretical predictions for soft helices. 
In the absence of pitch inversion, the pitch can be described by a simple algebraic density scaling  $P \propto \phi^{-1}$,  in good agreement with experimental observations in chiral nematics of filamentous virus particles and theoretical predictions. Irrespective of the patchyness of the rods, we find that the twist elastic constant of a nematic phase of repulsive spherocylinders grows linearly with packing fraction even in the proximity of the nematic-smectic transition, contrary to common belief where an anomalous growth or divergence is expected close to the transition \cite{gennes-prost}.

Future efforts could be aimed at  a careful measurement of orientational pair-correlation functions and effective pair-twisting potentials  with the aim to investigate whether the mechanism behind the cholesteric sense inversion reported in our simulation can somehow be connected to an effective  double-minimum twisting potential such as found for soft helices;  the minima are located at opposite signs of the intermolecular angle and suggest that i) helices with a certain molecular pitch have the propensity to twist in both directions and ii) the preferred twist direction is dictated by the amount of local orientational freedom the helices experience \cite{wensink2014}.  The data for the effective multi-particle torque density of the helices reported here point to a mechanism very similar to the one put forward in the theory.  Above all, our simulation results clearly demonstrate the robustness of these chirality inversions in twisted nematics of helical mesogens. They are not artefacts of the second-virial (and other  approximations) inherent to the theory. Nor are they suppressed by fluctuations in the director field or by density fluctuations generating local smectic order.

On a more ambitious level, it would be intriguing  to investigate to what extent kinetic factors contribute to the  handedness of the macroscopic pitch.
Recent simulation studies show that the chirality of fibrillar aggregates not only depends on the chirality of the constituent fibres
but also on the kinetic pathway of the self-assembly process \cite{gruziel2013}. A temporal change, for instance, in the local ionic strength may affect the binding sites between the filaments, thereby changing the molecular helicity which, in turn, may lead to chirality inversion on the fibre scale \cite{gruziel2015}.

Short-fragment DNA  strands may polymerize into longer units which subsequently self-organize into cholesteric phases.
Recent attempts have been made \cite{demichele2016} to model these systems from a theoretical  (DFT) perspective, focussing solely on shape chirality. The challenge here would be to include electrostatic chirality (on a coarse-grained level as done here) and see how both effects combined influence the pitch sensitivity and, indeed, the stability of the cholesteric state itself. In case of pronounced shape chirality, other chiral nematic phases different from the cholesteric may prevail and pair correlation functions measuring correlations between {\em e.g.} the azimuthal helix orientations could shed light on the stability of so-called twist-bend \cite{dozovEPL2001, borschNC2013} or screw-like nematics  \cite{kolli2014a} which may emerge in these systems. 
Last but not least, particle flexibility is believed to play an important role in the cholesteric self-organization of DNA and {\em fd} virus rods but its precise implications remain obscure and need to be addressed from a theoretical perspective.

\section{Acknowledgments}

The authors would like to thank  Michael P. Allen for helping us with implementing the simulation methodology. Tanja Schilling, Eric Grelet and Juho Lintuvuori are gratefully acknowledged for stimulating discussions. The project was funded by LabEX PALM (project {\em UPSCALE}).

\footnotesize{
\bibliography{rsc} 
\bibliographystyle{rsc} 
}

\end{document}